\documentstyle[11pt,cosmos,epsfig]{article}

\begin{document}

\begin{flushright}
FISIST/03-2001/CFIF
\end{flushright}
\title{Cosmological Constraints in 
SUSY with Yukawa Unification}

\author{Mario E. G\'omez}
\affil{Centro de F\'{\i}sica das 
Interac\c{c}\~{o}es Fundamentais (CFIF),  
Departamento de F\'{\i}sica, Instituto Superior T\'{e}cnico, 
Av. Rovisco Pais, 1049-001 Lisboa, Portugal}

\begin{abstract}
The cosmological relic density of the lightest 
supersymmetric particle of the minimal supersymmetric 
standard model is calculated under the assumption of 
gauge and Yukawa coupling unification. We employ radiative 
electroweak breaking with universal boundary conditions 
from gravity-mediated supersymmetry breaking. Further constraints are 
imposed by the experimental bounds on the $b$--quark mass and the
$BR(b\rightarrow s \gamma)$. We find that coannihilation 
of the lightest supersymmetric particle, which turns out to 
be an almost pure bino, with the next-to-lightest 
supersymmetric particle (the lightest stau) is crucial 
for reducing its relic density to an acceptable level.  
\end{abstract}

\keywords{Cosmology}

\section{Introduction}
In recent years the consideration of exotic dark matter has become necessary
in order to close the Universe [\cite{Jungm}]. 
In the currently favored supersymmetric (SUSY)
extensions of the standard model, the most natural WIMP  and candidate 
for CDM is the LSP,
i.e. the lightest supersymmetric particle. In the most favored scenarios the
LSP is the lightest neutralino, which can be simply described as a 
Majorana fermion, a linear 
combination of the neutral components of the gauginos and Higgsinos
[\cite{Jungm,haim}]. Its stability is guaranteed 
by imposing R-parity conservation, which implies that the LSP's 
can disappear only by annihilating in pairs.
\par
The simplest and most restrictive version of MSSM with gauge 
coupling unification is based on the assumption of radiative 
electroweak symmetry breaking with universal boundary conditions 
from gravity-mediated soft SUSY breaking. An interesting  
question is whether this scheme is compatible with exact 
``asymptotic" unification of the three third family Yukawa 
couplings. A positive answer to this question would be very 
desirable since it would lead to a simple and highly predictive 
theory. 

In this presentation we summarize the results of Refs.~[\cite{cdm,cdm2}],
aimed to answer the question of whether the simple version of the MSSM
with Yukawa unification at the GUT scale,
can be compatible with the most restrictive phenomenological constrains,
which are the correct predictions for $b$--quark mass and 
$BR(b\rightarrow s \gamma)$, and satisfy
the requirement that the 
relic abundance $\Omega_{LSP}~h^2$ of the lightest 
supersymmetric particle (LSP) in the universe does not exceed 
the upper limit on the cold dark matter (CDM) abundance implied 
by cosmological considerations.   

\section{Input parameters on the MSSM with Yukawa Unification}

We consider the MSSM embedded in some general supersymmetric 
GUT based on a gauge group such as $SO(10)$ or $E_{6}$ 
(where all the particles of one family belong to a single 
representation) with the additional requirement that the top, 
bottom and tau Yukawa couplings unify [\cite{als}] at the 
GUT scale $M_{GUT}$. Ignoring the Yukawa 
couplings of the first and second generation, the effective 
superpotential below $M_{GUT}$ is 
\begin{equation}
W=\epsilon_{ij}(-h_t H_2^i Q_3^j t^c+ h_b H_1^i Q_3^j b^c 
+ h_\tau H_1^i L_3^j \tau^c + \mu H_1^i H_2^j)~,
\label{super}
\end{equation}
where $Q_3=(t,b)$ and $L_3=(\nu_{\tau},\tau)$ are 
the quark and lepton $SU(2)_{L}$ doublet left handed 
superfields of the third generation and $t^c$, $b^c$ 
and $\tau^c$ the corresponding $SU(2)_{L}$ singlets. 
Also, $H_1$, $H_2$ are the electroweak higgs superfields 
and $\epsilon_{12}=+1$. The gravity-mediated soft 
supersymmetry breaking terms in the scalar potential are 
given by 
$$    
V_{soft}= \sum_{a,b} m_{ab}^{2}\phi^{*}_a \phi_b+ 
$$
\begin {equation}   
\left(\epsilon_{ij}(-A_t h_t  H_2^i \tilde Q_3^j \tilde t^c
+A_b h_b H_1^i \tilde Q_3^j\tilde b^c 
+A_\tau h_\tau H_1^i \tilde L_3^j \tilde\tau^c
+ B\mu H_1^iH_2^j)+ h.c.\right)~,
\label{vsoft}
\end{equation}
where the $\phi_a$ 's are the (complex) scalar fields and 
tildes denote superpartners. The gaugino mass terms in the 
Lagrangian are 
\begin{equation}
\frac{1}{2}(M_1 \tilde B\tilde B + 
M_2 \sum_{r=1}^{3} \tilde W_r \tilde W_r +  
M_3 \sum_{a=1}^{8}\tilde g_a\tilde g_a+h.c.)~,
\label{gaugino}
\end{equation}
where $\tilde B$, $\tilde W_r$ and $\tilde g_a$ are 
the bino, winos and gluinos respectively. `Asymptotic' 
Yukawa coupling unification implies
\begin{equation}
h_t(M_{GUT})=h_b(M_{GUT})=h_{\tau}(M_{GUT})\equiv h_0~.
\label{yukawa}
\end{equation}
Based on $N=1$ supergravity, we take universal soft 
supersymmetry breaking terms at $M_{GUT}$, i.e., a 
common mass for the scalar fields $m_0$, a common 
trilinear scalar coupling $A_0$ and $B_0=A_0-m_0$. 
Also, a common gaugino mass $M_{1/2}$ is assumed at 
$M_{GUT}$.
\par
Our effective theory below $M_{GUT}$ depends on the 
parameters ($\mu_0=\mu(M_{GUT})$)
\[
m_0,\ M_{1/2},\ A_0,\ \mu_0,\ \alpha_G,\ M_{GUT},
\ h_{0},\ \tan\beta~.  
\]
\begin{description}
\item[$\alpha_G$] and $M_{GUT}$  are evaluated 
consistently with the experimental values of $\alpha_{em},\ 
\alpha_s$ and $\sin^2\theta_W$ at $m_Z$. We integrate 
numerically the renormalization group equations (RGEs) 
for the MSSM at two loops in the gauge and Yukawa couplings 
from $M_{GUT}$ down to a common supersymmetry threshold 
$M_S=\sqrt{m_{\tilde t_1}
m_{\tilde t_2}}$. From this energy to $m_Z$, the 
RGEs of the nonsupersymmetric standard model are used.
\item[$\tan\beta$] is estimated at the scale  $M_S$
using the experimental input $m_\tau(m_{\tau})=1.777~\rm{GeV}$. 
We incorporate 
the SUSY threshold correction to $m_\tau(M_S)$ from the 
approximate formula of Ref.[\cite{pierce}]. It is about $8\%$, 
for $\mu>0$, leading to a value of 
$\tan\beta=55.4-54.5$ for $m_A=100-700~{\rm{GeV}}$, while, 
for $\mu<0$, we find a correction of about $-7\%$ and 
$\tan\beta=47.8-46.9$ in the same range of $m_A$. 
\item[$h_{0}$] is found by fixing the top quark mass at 
the center of its experimental range, 
$m_t(m_t)=166~{\rm{GeV}}$. The value obtained for $m_b(m_Z)$ after 
including supersymmetric corrections is somewhat higher 
than the experimental limit.
\item[$A_0$], for simplicity we take $A_0=0$. Our results 
move very little for negative values of $A_0$ bigger than 
about $-.5 M_{1/2}$, however lower negative values and positive 
values of this parameter tend to increase the SUSY spectrum increasing 
$\Omega_{LSP}~h^2$. Therefore the limits on  $\Omega_{LSP}~h^2$
imposes lower and upper bounds on $A_0$.
\item[$m_0,\ M_{1/2}, \mu_0,$] As we will discuss later the electroweak
symmetry breaking, when Yukawa unification is assumed, imposes a relation
on the values of $m_0$ and $M_{1/2}$. On the other hand the role of 
coannihilation $\tilde\chi$--$\tilde\tau$ make convenient to express 
our results in terms of relative mass splitting  $\Delta_{\tilde\tau_2}
=(m_{\tilde\tau_2}-m_{\tilde\chi})/m_{\tilde\chi}$ 
between the NLSP and LSP. Therefore we trade the GUT values of
$m_0,\ M_{1/2}$ and  $\mu_0$ by the pseudoscalar Higgs mass $m_A$ 
and $\Delta_{\tilde\tau_2}$.
\end{description}
Let us describe with more detail the last item above. Assuming 
radiative electroweak 
symmetry breaking, we can express the values of the parameters 
$\mu$ (up to its sign) and $B$ at $M_S$ in terms of the 
other input parameters by means of the appropriate conditions
\begin{equation}
\mu^2=\frac{m^2_{H_1}-m^2_{H_2}\tan^2{\beta}}
{\tan^2{\beta}-1}- \frac{1}{2} m^2_Z \ , \ \sin 2\beta
=-\frac{2 B \mu}{m_{H_1}^2+m_{H_2}^2+2 \mu^2}~,
\label{mu}
\end{equation}
where $m_{H_1}$, $m_{H_2}$ are the soft 
supersymmetry breaking scalar higgs masses. When unified Yukawa couplings
and a common value for $m_{H_2}$ and  $m_{H_1}$ 
are assumed  GUT , we find both  $m_{H_2}^2$ and  $m_{H_1}^2$ negative
at $M_S$. However for certain values of $m_0$ and $M_{1/2}$ is possible
to find values the pseudoscalar Higgs,
\[
m_A=m_{H_1}^2+m_{H_2}^2+2 \mu^2
\]
beyond a lower bound, considered to be $m_Z$ in the present work. 
Furthermore, the authors of Ref.~[\cite{anant}] found that, for every 
value of $m_A$ and a fixed value of $m_t(m_t)$, there 
is a pair of minimal values of $m_0$ and $M_{1/2}$ 
where the masses of the LSP and $\tilde\tau_2$ 
are equal. This is understood from 
the dependence of $m_A$ on $m_0$ and $M_{1/2}$ given in 
Ref.~[\cite{copw}]:
\begin{equation}
m^2_A = \alpha M^2_{1/2}-\beta m^2_0-{\rm{const.}}~,
\label{mAc}
\end{equation} 
where all the coefficients are positive and $\alpha$ and 
$\beta$, which depend only on $m_t(m_t)$, are $\sim 0.1$
(the constant turns out to be numerically close to $m_Z^2$). 
Equating the masses of the LSP and 
$\tilde\tau_2$ is equivalent to relating $m_0$ and 
$M_{1/2}$. Then, for every $m_A$, a pair of values of 
$m_0$ and $M_{1/2}$ is determined. 
We had included the full one--loop radiative corrections to the 
effective potential as given in the appendix E of Ref.~[\cite{pierce}].
 
The values of
the LSP, $M_S$ and the corresponding values of $m_0$ and $M_{1/2}$
are given in Fig.1.

\begin{figure}
\centerline{\epsfig{figure=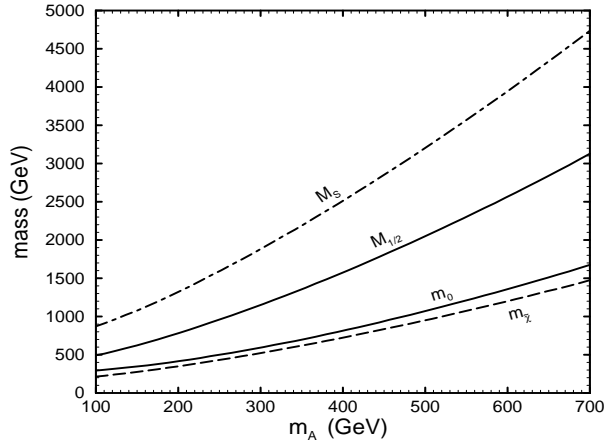,width=8.cm,height=6cm}}
\medskip
\caption{The values of $m_{\tilde\chi}$, $m_0$, $M_{1/2}$ 
and $M_S$ as functions of $m_A$ for $\mu>0$, $A_0=0$ and 
$m_{\tilde\tau_2}=m_{\tilde\chi}$. These values are affected 
very little by changing the sign of $\mu$.
\label{ms}}
\end{figure}
\section{Phenomenological constraints from $m_b$ and $b\rightarrow s \gamma$}
A significant problem, which may be faced in trying to reconcile 
Yukawa unification and universal boundary conditions, is due to 
the generation of sizeable SUSY corrections to the $b$-quark 
mass [\cite{copw,hall}]. The sign of these corrections is 
opposite to the sign of the MSSM parameter $\mu$ (with the 
conventions of Ref.[\cite{cdm}]). As a consequence, for $\mu<0$, 
the tree-level value of $m_b$, which is 
predicted from Yukawa unification already near its experimental 
upper bound, receives large positive corrections which drive 
it well outside the allowed range. However, it  
should be noted that this problem arises in the simplest 
realization of this scheme. In complete models correctly
incorporating fermion masses and mixing, $m_b$ can receive
extra corrections which may make it compatible with experiment. 
So, we do not consider this $b$-quark 
mass problem absolutely fatal for the $\mu<0$ case.
However in the alternative scenario, with $\mu>0$, the $b$-quark mass 
receives negative SUSY corrections and can easily be compatible 
with data in this case. An example of the typical values we find
is given in table I.

\begin{figure}
\centerline{\epsfig{figure=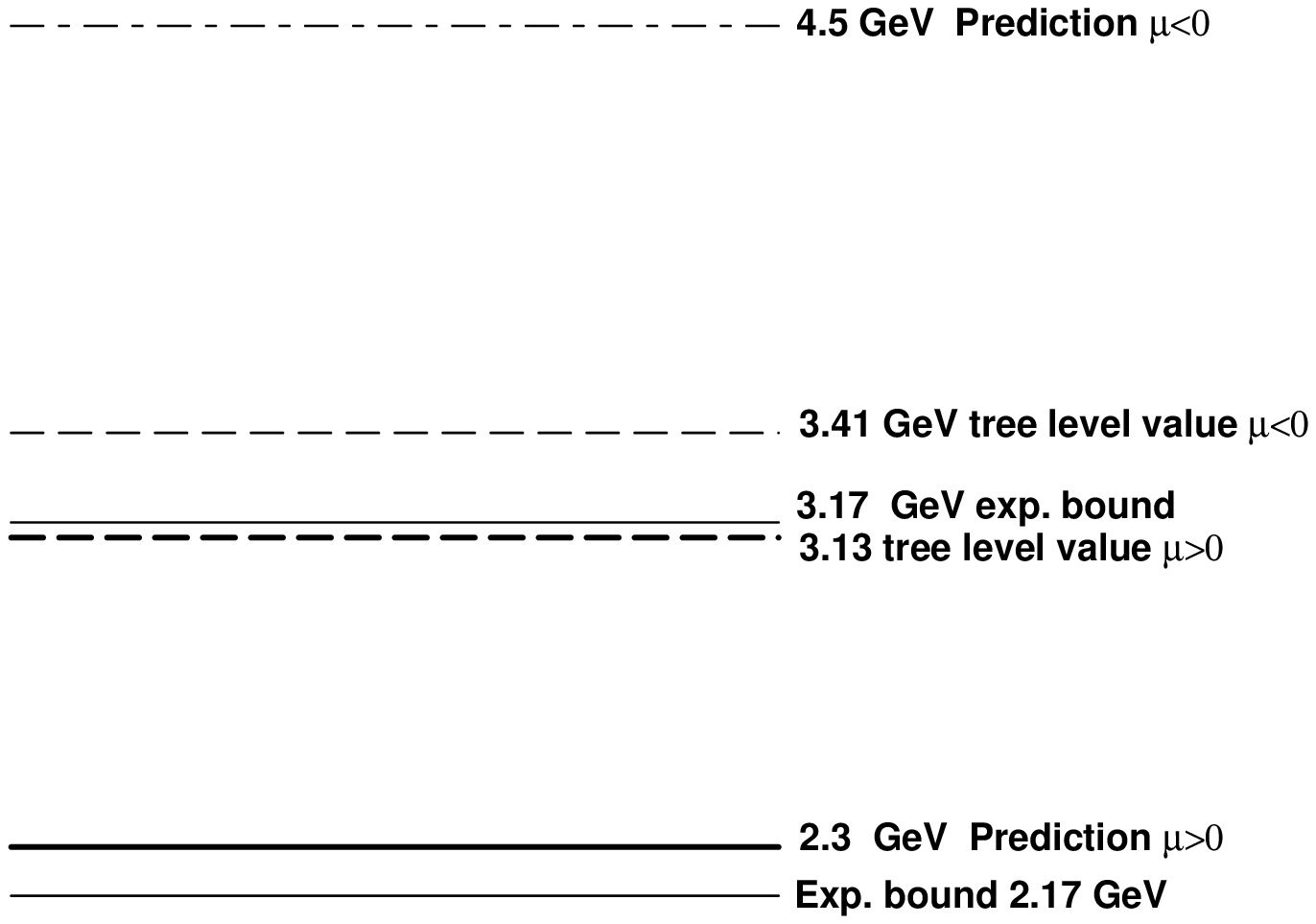,width=6.cm,height=4cm}}
\medskip
\begin{center}
Table I. Values for the quark bottom mass.
\end{center}
\end{figure}
\par

This scheme with $\mu>0$,  is severely
restricted by the recent experimental results [\cite{cleo}] on 
the inclusive decay $b\rightarrow s\gamma$ [\cite{bsg}]. It 
is well-known that the SUSY corrections to the inclusive 
branching ratio BR($b\rightarrow s\gamma$), in the case of 
the MSSM with universal boundary conditions, arise mainly 
from chargino loops and have the same 
sign with the parameter $\mu$. Consequently, these 
corrections interfere constructively with the contribution 
from the standard model (SM) including an extra electroweak 
Higgs doublet. However, this contribution is already bigger than 
the experimental upper bound on BR($b\rightarrow s\gamma$) 
for not too large values of the CP-odd Higgs boson mass $m_A$. 
As a result, in the present context with Yukawa unification and 
hence large $\tan\beta$, a lower bound on $m_A$ is obtained 
for $\mu>0$. On the contrary, for 
$\mu<0$, the SUSY corrections to BR($b\rightarrow s\gamma$)
interfere destructively with the SM plus extra Higgs doublet 
contribution yielding, in most cases, no restrictions on the 
parameters. The results corresponding to the parameters 
given in fig.1  are shown in fig.2. In the case of 
$\mu>0$, BR($b\rightarrow s\gamma$) decreases as the splitting between 
$m_{\tilde\tau_2}$ and $m_{\tilde\chi}$ increases.

\begin{figure}
\centerline{\epsfig{figure=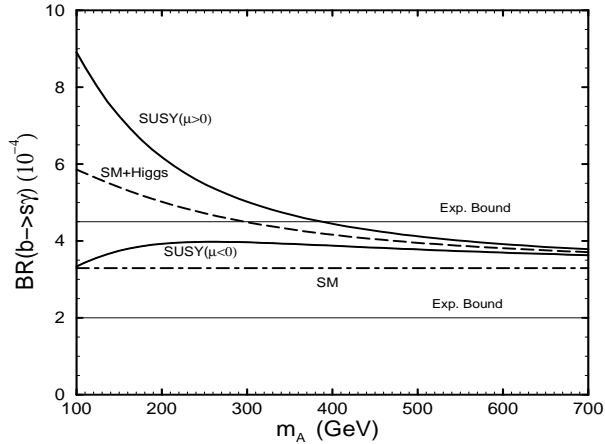,width=8.cm,height=6cm}}
\medskip
\caption{The central value of the SUSY inclusive 
BR($b\rightarrow s\gamma$) as function of $m_A$ for both signs 
of $\mu$, $A_0=0$ and $m_{\tilde\tau_2}=m_{\tilde\chi}$. 
The contributions from the SM and the SM plus charged Higgs boson 
(SM+Higgs) as well as the experimental bounds on 
BR($b\rightarrow s\gamma$), $2\times 10^{-4}$ and 
$4.5\times 10^{-4}$, are also indicated.    
\label{bsg}}
\end{figure}

\section{LSP relic abundance and bino--stau coannihilations }
The cosmological constraint on the parameter space 
results from the requirement that the 
relic abundance $\Omega_{LSP}~h^2$ of the lightest 
supersymmetric particle (LSP) in the universe does not exceed 
the upper limit on the cold dark matter (CDM) abundance implied 
by cosmological considerations ($\Omega_{LSP}$ is the present 
energy density of the LSPs over the critical energy density of 
the universe and $h$ is the present value of the Hubble constant 
in units of $100~\rm{km}~\rm{sec}^{-1}~\rm{Mpc}^{-1}$). Taking 
both the currently available cosmological models with 
zero/nonzero cosmological constant, which provide the best fits 
to all the data, as equally plausible alternatives for the 
composition of the energy density of the universe and accounting 
for the observational uncertainties, we obtain the restriction 
$\Omega_{LSP}~h^{2}\stackrel{_{<}}{_{\sim }}0.22$ (see 
Refs.[\cite{cdm})]. Assuming that all the CDM in the 
universe is composed of LSPs, we further get 
$\Omega_{LSP}~h^{2}\stackrel{_{>}}{_{\sim }}0.09$.

The cosmological relic 
density of the lightest neutralino $\tilde\chi$ (almost 
pure $\tilde B$) in MSSM with Yukawa unification increases to 
unacceptably 
high values as $m_{\tilde\chi}$ becomes larger. 
Low values of $m_{\tilde\chi}$ are obtained when 
the NLSP ($\tilde\tau_2$) is almost degenerate with 
$\tilde\chi$. Under these circumstances, 
coannihilation of $\tilde\chi$ with $\tilde\tau_2$ and 
$\tilde\tau_2^\ast$ is of crucial importance 
reducing further the $\tilde\chi$ relic density by a 
significant amount. The important role of coannihilation of 
the LSP with sparticles carrying masses close to its mass in 
the calculation of the LSP relic density has been pointed 
out by many authors (see e.g., Refs.[\cite{drees,ellis,coan})]. 
Here, we will use the method described by Griest and Seckel 
[\cite{coan}]. Note that our analysis can be readily applied 
to any MSSM scheme where the LSP and NLSP are the bino and 
stau respectively.

The relic abundance of the LSP at the present cosmic time 
can be calculated from the equation: 
\begin{equation}
\Omega_{\tilde\chi}~h^2\approx\frac{1.07 \times 10^9 
~{\rm GeV}^{-1}}{g_*^{1/2}M_{P}~x_F^{-1}~
\hat\sigma_{eff}}
\label{omega}
\end{equation}
with
\begin{equation}
\hat\sigma_{eff}\equiv x_F\int_{x_F}^{\infty}
\langle\sigma_{eff}v\rangle x^{-2}dx~.
\label{sigmaeff3}
\end{equation}
Here $M_P=1.22 \times 10^{19}$ GeV is the Planck scale, 
$g_*\approx 81$ is the effective number of massless 
degrees of freedom at freeze-out and 
$x_F=m_{\tilde\chi}/T_{F}$, with $T_F$ being the 
freeze-out photon temperature.

In our case, $\sigma_{eff}$ takes the form 
\begin{equation}
\sigma_{eff}= 
\sigma_{\tilde\chi\tilde\chi}
r_{\tilde\chi}r_{\tilde\chi}+
4\sigma_{\tilde\chi\tilde\tau_2}
r_{\tilde\chi}r_{\tilde\tau_2}+
2(\sigma_{\tilde\tau_2\tilde\tau_2}+
\sigma_{\tilde\tau_2\tilde\tau_2^\ast})
r_{\tilde\tau_2}r_{\tilde\tau_2}~.
\label{sigmaeff2}
\end{equation}
For $r_i$, we use the nonrelativistic approximation
\begin{equation}
r_i(x) = \frac{g_i (1+\Delta_i)^{3/2} e^{-\Delta_i x}}
{g_{eff}}~,
\label{ri}
\end{equation}
\begin{equation}
g_{eff}(x)={\sum_{i}g_i (1+\Delta_i)^{3/2} 
e^{-\Delta_i x}},
~\Delta_i=(m_i-m_{\tilde\chi})/m_{\tilde\chi}~.
\label{geff}
\end{equation}
Here $g_i=2$, 1, 1 ($i=\tilde\chi$, $\tilde\tau_2$, 
$\tilde\tau_2^\ast$) is the number of degrees of 
freedom of the particle species $i$ with mass $m_i$ and 
$x=m_{\tilde\chi}/T$ with $T$ being the photon 
temperature.

The freeze-out temperatures which we obtain here are of the 
order of $m_{\tilde\chi}/25$ and, thus, our nonrelativistic 
approximation (see Eq.(\ref{ri})) is justified. Under 
these circumstances, the quantities $\sigma_{ij}v$ are well 
approximated by their Taylor expansion up to second order in 
the `relative velocity', 
\begin{equation}
\sigma_{ij}v=a_{ij}+b_{ij} v^2~.
\label{taylorv}
\end{equation}
The thermally averaged cross sections are then easily 
calculated 
\begin{equation}
\langle \sigma_{ij} v \rangle (x)= 
\frac{x^{3/2}}{2 \sqrt{\pi}} 
\int_{0}^{\infty} d v v^2 (\sigma_{ij} v) e^{-x v^2/4}
=a_{ij}+6 b_{ij}/x~.
\label{average}
\end{equation}

The contribution of the $a_{ij}$'s  to the cross section
 is more important than the  than the  $b_{ij}$'s since its contribution
is suppressed by a factor $6/x_F\approx .2$. The annihilation
cross section is suppressed respect the coannihilation channels due to the
fact that $a_{\tilde\chi\tilde\chi }$ is suppressed due to Fermi statistics
[\cite{haim}]. This suppression is not present in the coannihilation channels,
however its contribution to $\sigma_{eff}$ is attenuated by the 
exponential in $r_{\tilde{\tau}_2}$ in eq.~(10). Therefore 
coannihilations have an 
important effect in decreasing the $\Omega_{LSP}$ when the LSP and 
the NLSP are very close in mass. In our case  we find 
this effect negligible when the mass splitting 
$\tilde{\chi}$--$\tilde{\tau}$ is greater than approximately 20\%. 
The complete list of Feynman diagrams and the expressions for the
$a_{ij}$'s appropriate for large $\tan\beta$ are given in Ref.~[\cite{cdm}].

The result of including coannihilation of $\tilde\tau_2$ and $\tilde\chi$
in the computation of the LSP relic abundance is clearly shown in fig.3.
The effect of increase the mass splitting between  
$\tilde\tau_2$ and $\tilde\chi$ will result in larger values for 
$\Omega_{LSP}$.

\begin{figure}
\centerline{\epsfig{figure=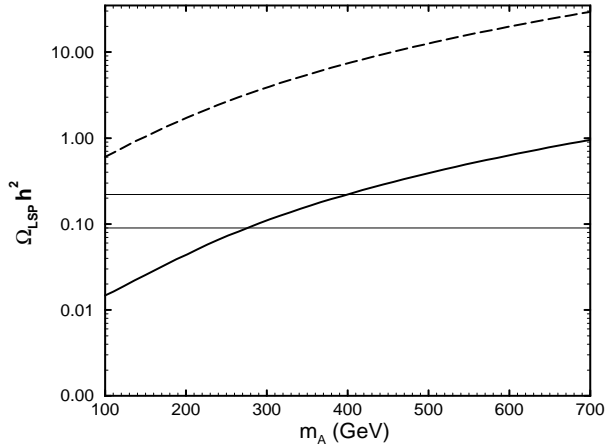,width=8.cm,height=6cm}}
\medskip
\caption{The LSP relic abundance $\Omega_{LSP}~h^2$ as function of 
$m_A$ in the limiting case $m_{\tilde\tau_2}=m_{\tilde\chi}$ and 
for $\mu>0$, $A_0=0$. The solid line includes coannihilation of 
$\tilde\tau_2$ and $\tilde\chi$, while the dashed line is obtained 
by only considering the LSP annihilation processes. These results are 
affected very little by changing the sign of $\mu$. The limiting lines 
at $\Omega_{LSP}~h^2=0.09$ and 0.22 are also included.
\label{lspp}}
\end{figure}

\section{Conclusions}

We have shown that the condition of ``asymptotic" Yukawa unification on
the MSSM results in a significant constraint on the free parameter 
space of the model.

Constraints from $m_b$ and $b\rightarrow s \gamma$ can be 
simultaneously satisfied for $\mu>0$ and relatively large values of 
the SUSY parameters $m_A\sim 385 {\rm GeV}$, 
$m_{\tilde{\chi}}\sim 695{\rm GeV}$, $m_0\sim 780 {\rm GeV}$ and
$M_{1/2}\sim 1.5 {\rm TeV}$. The constraint derived from cosmological
limits on LSP relic abundance, is satisfied only when 
bino--stau coannihilations are relevant. If we superpose figs. 1, 2, 3 we can 
observe that the previous conditions are satisfied in a narrow band of
parameter space.

The hight values of $m_{\tilde{\chi}}$ in the parameter space allowed
by the constraints described above will make difficult its direct 
detection. However the large $\tan\beta$ scheme can provide interesting 
predictions [\cite{det}] if one relaxes the strict 
unification condition imposed in the present study.

\acknowledgments
This work was supported by the European Union under TMR contract 
No. ERBFMRX--CT96--0090.
\def\ijmp#1#2#3{{ Int. Jour. Mod. Phys. }{\bf #1~}(#2)~#3}
\def\pl#1#2#3{{ Phys. Lett. }{\bf B#1~}(#2)~#3}
\def\zp#1#2#3{{ Z. Phys. }{\bf C#1~}(#2)~#3}
\def\prl#1#2#3{{ Phys. Rev. Lett. }{\bf #1~}(#2)~#3}
\def\rmp#1#2#3{{ Rev. Mod. Phys. }{\bf #1~}(#2)~#3}
\def\prep#1#2#3{{ Phys. Rep. }{\bf #1~}(#2)~#3}
\def\pr#1#2#3{{ Phys. Rev. }{\bf D#1~}(#2)~#3}
\def\np#1#2#3{{ Nucl. Phys. }{\bf B#1~}(#2)~#3}
\def\npps#1#2#3{{ Nucl. Phys. (Proc. Sup.) }{\bf B#1~}(#2)~#3}
\def\mpl#1#2#3{{ Mod. Phys. Lett. }{\bf #1~}(#2)~#3}
\def\arnps#1#2#3{{ Annu. Rev. Nucl. Part. Sci. }{\bf
#1~}(#2)~#3}
\def\sjnp#1#2#3{{ Sov. J. Nucl. Phys. }{\bf #1~}(#2)~#3}
\def\jetp#1#2#3{{ JETP Lett. }{\bf #1~}(#2)~#3}
\def\app#1#2#3{{ Acta Phys. Polon. }{\bf #1~}(#2)~#3}
\def\rnc#1#2#3{{ Riv. Nuovo Cim. }{\bf #1~}(#2)~#3}
\def\ap#1#2#3{{ Ann. Phys. }{\bf #1~}(#2)~#3}
\def\ptp#1#2#3{{ Prog. Theor. Phys. }{\bf #1~}(#2)~#3}
\def\plb#1#2#3{{ Phys. Lett. }{\bf#1B~}(#2)~#3}
\def\apjl#1#2#3{{ Astrophys. J. Lett. }{\bf #1~}(#2)~#3}
\def\n#1#2#3{{ Nature }{\bf #1~}(#2)~#3}
\def\apj#1#2#3{{ Astrophys. Journal }{\bf #1~}(#2)~#3}
\def\anj#1#2#3{{ Astron. J. }{\bf #1~}(#2)~#3}
\def\mnras#1#2#3{{ MNRAS }{\bf #1~}(#2)~#3}
\def\grg#1#2#3{{ Gen. Rel. Grav. }{\bf #1~}(#2)~#3}
\def\s#1#2#3{{ Science }{\bf #1~}(19#2)~#3}
\def\baas#1#2#3{{ Bull. Am. Astron. Soc. }{\bf #1~}(#2)~#3}
\def\ibid#1#2#3{{ ibid. }{\bf #1~}(19#2)~#3}
\def\cpc#1#2#3{{ Comput. Phys. Commun. }{\bf #1~}(#2)~#3}
\def\astp#1#2#3{{ Astropart. Phys. }{\bf #1~}(#2)~#3}
\def\epj#1#2#3{{ Eur. Phys. J. }{\bf C#1~}(#2)~#3}

\thebibliography

\bibitem{Jungm}[1]For a recent review see e.g.
G. Jungman {\it et al.},{\it  Phys. Rep.} {\bf 267}, 195  (1996).
\bibitem{als}[2] B. Ananthanarayan, G. Lazarides and Q. Shafi,
\pr{44}{1991}{1613}.
\bibitem{cdm}[3] M. G\'{o}mez, G. Lazarides and C. Pallis, 
\pr{61}{2000}{123512}.
\bibitem{cdm2}[4] M. G\'{o}mez, G. Lazarides and C. Pallis, 
\plb{487}{2000}{313}.
\bibitem{copw}[5] M. Carena, M. Olechowski,
S. Pokorski and C. E. M. Wagner, \np{426}{1994}{269}.
\bibitem{anant}[6] B. Ananthanarayan, Q. Shafi and X. M. Wang, 
\pr{50}{1994}{5980}.
\bibitem{hall}[7] L. Hall, R. Rattazzi and U. Sarid,
\pr{50}{1994}{7048};
R. Hempfling, \pr{49}{1994}{6168}.
\bibitem{cleo}[8] CLEO Collaboration (S. Glenn {\it{et al.}}), 
CLEO CONF 98-17, talk presented at the XXIX ICHEP98, UBC, Vancouver, 
B. C., Canada, July 23-29 1998;
ALEPH Collaboration (R. Barate {\it{et al.}}),\pl{429}{1998}{169}.

\bibitem{bsg}[9] S. Bertolini, F. Borzumati, A. Masiero and G. Ridolfi, 
\np{353}{1991}{591}; 
\bibitem{haim}[10]H. Goldberg, {\it Phys. Rev. Lett} {\bf  50}, 1419 (1983);

\bibitem{coan}[11] K. Griest and D. Seckel, \pr{43}{1991}{3191}.

\bibitem{drees}[12] M. Drees and M. M. Nojiri, 
\pr{47}{1993}{376}; 
S. Mizuta and M. Yamaguchi, \pl{298}{1993}{120}; 
P. Gondolo and J. Edsj\"o, \pr{56}{1997}{1879};

\bibitem{ellis}[13] J. Ellis, T. Falk, K. A. Olive and M. Srednicki, 
\astp{13}{2000}{181}.

\bibitem{pierce}[14] D. Pierce, J. Bagger, K. Matchev and R. Zhang,
\np{491}{1997}{3}.

\bibitem{det}[15] M.E. G\'omez and J.D. Vergados, hep-ph/0012020. 

\endthebibliography

\end{document}